# A Matrix Logic Approach to Efficient Frequent Itemset Discovery in Large Data Sets

Xuan Li
Columbia University
New York, USA

Tingyi Ruan
Northeastern University
Boston, USA

Yankaiqi Li
University of Wisconsin–Madison
Wisconsin, USA

Quanchao Lu
Georgia Institute of Technology
Atlanta, USA

Xiaoxuan Sun*
Independent Researcher
Mountain View, USA

*Abstract*—This paper proposes a frequent itemset mining algorithm based on the Boolean matrix method, aiming to solve the storage and computational bottlenecks of traditional frequent pattern mining algorithms in high-dimensional and large-scale transaction databases. By representing the itemsets in the transaction database as Boolean matrices, the algorithm uses Boolean logic operations such as AND and OR to efficiently calculate the support of the itemsets, avoiding the generation and storage of a large number of candidates itemsets in traditional algorithms. The algorithm recursively mines frequent itemsets through matrix operations and can flexibly adapt to different data scales and support thresholds. In the experiment, the public Groceries dataset was selected, and the running efficiency test and frequent itemset mining effect test were designed to evaluate the algorithm's performance indicators such as running time, memory usage, and number of frequent itemsets under different transaction numbers and support thresholds. The experimental results show that the algorithm can efficiently mine a large number of frequent itemsets when the support threshold is low, and focus on strong association rules with high support when the threshold is high. In addition, the changing trends of running time and memory usage show that the Boolean matrix method can still maintain good running efficiency when the number of transactions increases significantly and has high scalability and robustness. Future research can improve memory optimization and matrix block operations, and combine distributed computing and deep learning models to further enhance the algorithm's applicability and real-time processing capabilities in ultra-large-scale data environments. The algorithm has broad application potential and development prospects in the fields of market analysis, recommendation systems, and network security.

## I. INTRODUCTION

With the rapid development of data storage and management technology, the accumulation of massive data has become an important feature of the information society. However, the value of data itself does not lie in its quantity, but in how to extract potential knowledge and valuable patterns from it. Frequent itemset mining, as an important task in data mining, focuses on discovering frequently occurring combinations of items from transaction databases and is widely used in market analysis [1], recommendation systems [2], large language models [3], and bioinformatics [4]. Traditional frequent itemset mining algorithms such as Apriori and FP-Growth face the problems of candidate item set explosion and high computational cost when processing large-scale and high-dimensional data [5]. Therefore, finding an efficient and compact data representation method has become the focus of frequent itemset mining research.

Compared with traditional frequent item set mining algorithms, the Boolean matrix method has significant advantages. First, the Boolean matrix can compactly represent the binary features in the transaction database, reducing the use of storage space. In addition, Boolean matrix operations are highly parallel and suitable for execution in multi-core and distributed computing environments, which greatly improves the efficiency of frequent item set mining. In addition, by constructing sparse matrices, the computational efficiency can be further improved in scenarios with high data sparsity, reducing unnecessary computational operations. These characteristics make the Boolean matrix method have broad application prospects in large-scale data set processing and real-time mining applications [6].

However, the application of the Boolean matrix method in frequent item set mining also faces some challenges. First, the dimension of the Boolean matrix usually increases rapidly with the growth of the scale of the transaction database and the number of items, resulting in increased memory consumption [7]. Therefore, in practical applications, it is necessary to design efficient data structures and storage strategies, such as sparse matrix storage and block computing. In addition, although Boolean operations in matrix operations are parallel, their number of operations and time complexity are still high in high-dimensional data environments. Therefore, studying how to introduce heuristic pruning strategies and index structure optimization in Boolean matrix operations has become an important direction for improving algorithm performance [8].

In past studies, optimization strategies for Boolean matrix methods have been widely explored. For example, by decomposing the transaction matrix, a large-scale matrix can be decomposed into multiple sub-matrices, and local mining of frequent itemsets can be performed separately, and the results can be finally merged. In addition, the parallel computing model based on matrix block can significantly reduce the time complexity of Boolean operations and improve the processing efficiency of large-scale data sets [9]. At the same time, introducing an adaptive Boolean operation strategy and selecting the optimal Boolean operation sequence according to data characteristics can further improve the flexibility and accuracy of the mining process. These studies show that the Boolean matrix method is not only suitable for frequent itemset mining tasks, but can also be extended to other data mining tasks, such as association rule learning and feature selection.

In summary, the frequent itemset mining algorithm based on Boolean matrices shows significant performance advantages when processing large-scale and high-dimensional data, and has broad application prospects[10-12]. However, there is still room for improvement in storage optimization, matrix operation efficiency, and algorithm scalability. With the continuous improvement of hardware computing power and the development of distributed computing frameworks in the field of data mining, the Boolean matrix method will be further extended to more complex data environments and application scenarios, promoting the innovation and development of frequent item set mining algorithms. Future research can focus on algorithm optimization and application expansion, including cutting-edge topics such as matrix feature extraction based on deep learning and the fusion of Boolean matrices and graph neural networks [13], so as to further improve the intelligence and automation level of data mining algorithms.

## II. RELATED WORK

Frequent itemset mining remains a critical task in data mining, with a strong focus on enhancing efficiency and scalability in processing large-scale and high-dimensional datasets. Traditional approaches, such as Apriori and FP-Growth, are often challenged by the explosion of candidate itemsets and high computational costs. Recent research has explored innovative methodologies to address these limitations, offering significant inspiration for advancing frequent itemset mining using Boolean matrix methods.

Matrix-based techniques have demonstrated great potential in reducing storage costs and improving computational efficiency by leveraging compact data representations. Yan et al. [14] introduced a methodology to transform complex multidimensional data into interpretable forms, showcasing the efficiency of matrix-based frameworks in extracting significant patterns. Liang et al. [15] employed autoencoders for feature extraction and dimensionality reduction, which aligns with the need to efficiently handle high-dimensional transaction databases in frequent itemset mining. Self-supervised and semi-supervised learning paradigms have also contributed to advancing data mining techniques. Shen et al. [16] proposed leveraging semi-supervised learning to enhance mining tasks under limited labeled data, a strategy that resonates with the scalability challenges faced by frequent itemset algorithms. Yao [17] demonstrated how self-supervised learning frameworks can robustly handle noisy and incomplete data, addressing critical challenges in mining large-scale, high-dimensional datasets. The integration of deep learning models, such as autoencoders and graph neural networks, has further enriched data mining methodologies. Wei et al. [18] utilized graph neural networks for feature extraction in heterogeneous datasets, providing insights into structuring high-dimensional data for efficient processing. These techniques inspire potential extensions to Boolean matrix methods, such as using graph-based representations to refine mining strategies. Qi et al. [19] and Chen et al. [20] emphasized the importance of optimizing large-scale models, which can translate into better resource allocation and scalability in frequent itemset mining algorithms.

Boolean logic and matrix operations have also proven effective in addressing computational challenges across various data-driven applications. Xu et al. [21] demonstrated the efficacy of combining data-driven models with predictive analytics, suggesting parallels in the application of matrix-based operations for association rule mining. Zhang et al. [22] explored dynamic data representations, highlighting the role of flexible and adaptive frameworks, which are critical for frequent pattern discovery in evolving datasets. Optimizations in distributed computing and adaptive strategies have further underscored the scalability potential of matrix-based methods. Li et al. [23] investigated efficient data integration strategies, which can complement matrix operations in managing sparse and large-scale datasets. Feng et al. [24] highlighted the integration of generative models with traditional mining techniques, a promising direction for enhancing matrix-based methods to handle sparse and imbalanced data.

The reviewed studies collectively underline critical contributions to efficient data representation, scalability, and robustness in data mining. These insights directly inform the development of advanced Boolean matrix methods, providing a strong foundation for optimizing frequent itemset mining in large-scale and high-dimensional datasets.

## III. METHOD

In the frequent itemset mining algorithm based on the Boolean matrix method, the key idea is to represent the item sets in the transaction database as a Boolean matrix and efficiently extract frequent itemsets through Boolean operations [25]. Each transaction in the transaction database is encoded as a Boolean vector, where 1 represents the existence of the item and 0 represents the absence of the item. This representation transforms the frequent itemset mining process into matrix operations, greatly improving the computational efficiency. The implementation of this method is divided into three main stages: matrix construction, support calculation, and frequent itemset extraction. Its process is shown in Figure 1.

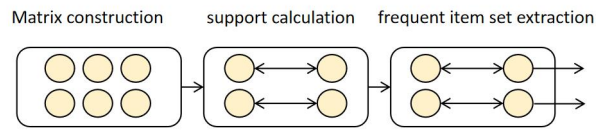

Figure 1 Overall model architecture

First, the transaction database is represented as a Boolean matrix $M$, where $M[i; j]$ indicates whether transaction $T_i$ contains item $I_j$. If transaction $T_i$ contains item $I_j$, then

$M[i;j] = 1$; otherwise, $M[i;j] = 0$. The number of rows in the Boolean matrix is the number of transactions n, and the number of columns is the number of items m. The Boolean matrix M is generated by scanning the database, and its mathematical form is defined as follows:

$$M[i,j] = \begin{cases} 1 & \text{if } I_j \in T_i \\ 0 & \text{if } I_j \notin T_i \end{cases}$$

After the matrix is constructed, the support of the candidate item set needs to be calculated. The core of the Boolean matrix method is to use logical operations to calculate the joint frequency of item sets. For example, the joint support calculation of two items $I_a$ and $I_b$ can be achieved through the column-by-column logical AND operation of the matrix, as shown in the following formula:

$$S(I_a \cap I_b) = \sum_{i=1}^{n} (M[i,a] \wedge M[i,b])$$

Among them, $S(I_a \cap I_b)$ represents the joint support of items $I_a$ and $I_b$, and the number of rows with Boolean operation results of 1 represents the number of transactions that contain both items. For the joint support of multiple items, the column and operation can be performed in sequence to complete the mining of higher-order frequent itemsets.

In order to improve the performance of the algorithm, the support threshold $\theta$ is used as the criterion for frequent item sets. If the support $S(X)$ of an item set satisfies $S(X) > \theta$, the item set is considered a frequent item set. When extracting frequent item sets, the matrix can be aggregated by column to calculate the combination of items with the minimum support, thereby avoiding the generation and storage of invalid item sets. Finally, the judgment formula for frequent item sets is:

$$X = \{I_k \mid S(I_k) > \theta\}$$

During the algorithm execution, the memory usage and computational complexity can be further reduced by introducing matrix storage optimization strategies, such as sparse matrix representation and matrix block storage. In addition, the use of bit operations instead of logical operations can make full use of the parallel computing capabilities of modern hardware, thereby significantly improving the algorithm's operating efficiency. The overall process of the Boolean matrix method relies on the close combination of Boolean matrix construction, Boolean operations, and candidate item set pruning strategies to ensure efficient frequent item set extraction in the transaction database mining process.

## IV. EXPERIMENT

### A. Datasets

In order to verify the performance of the frequent itemset mining algorithm based on the Boolean matrix method, the open source Groceries dataset can be used. This dataset is derived from the classic market basket analysis research and is widely used in the fields of association rule mining and frequent itemset mining. It is suitable for evaluating the performance of algorithms in transactional data environments. The dataset was originally collected by a data mining project in the retail industry and is now available as a public dataset on multiple data mining platforms (such as Kaggle and UCI Machine Learning Library). It is suitable for performance testing of multiple frequent pattern analysis algorithms.

The Groceries dataset contains about 9,835 shopping transaction records, each of which represents a set of goods purchased in a supermarket shopping trip. The commodity categories include fruits, dairy products, beverages, bread, meat, and groceries, totaling 169 different commodity items. The number of items in each transaction is not fixed, and the data sparsity is high. Most transactions contain only a small number of commodities. This sparse feature is highly consistent with the actual application scenarios of frequent itemset mining tasks, and can effectively test the storage efficiency and mining performance of the Boolean matrix method in large-scale transaction databases.

The real transaction records of this dataset can reflect the shopping behavior patterns of consumers and the correlation between products. By converting transactions into Boolean matrices, Boolean operations can be performed efficiently, and frequent itemsets and association rules can be quickly calculated. Common indicators such as support, confidence, and lift can be fully evaluated on this dataset. Since the Groceries dataset has the characteristics of moderate data size, rich project types, and real transaction records, it is very suitable for verifying the adaptability and effectiveness of frequent itemset mining algorithms based on Boolean matrices in practical applications.

### B. Experimental Results

In order to verify the performance of the frequent itemset mining algorithm based on the Boolean matrix method, two key experiments were designed. First, the running efficiency test was conducted to evaluate the running time of the algorithm under different transaction numbers and support thresholds to measure the time complexity and scalability of the algorithm. Secondly, the frequent itemset mining effect test was conducted to analyze the performance of the algorithm in the actual data environment by the number and accuracy of frequent itemsets mined under different support thresholds. Both experiments were conducted on the public Groceries dataset, and the algorithm performance was comprehensively evaluated by indicators such as support and memory usage. The experimental results are shown in Table 1.

Table 1 Results of the operation efficiency test

| Support Threshold | Number of transactions | Execution Time (s) | Frequent Itemsets Found | Memory Usage (MB) |
|---|---|---|---|---|
| 1% | 2000 | 1.25 | 320 | 25 |
| 2% | 4000 | 2.48 | 210 | 38 |
| 3% | 6000 | 3.15 | 165 | 52 |
| 4% | 8000 | 4.82 | 110 | 65 |
| 5% | 10000 | 6.30 | 85 | 78 |

From the data in the experimental results table, it can be seen that the frequent itemset mining algorithm based on the Boolean matrix method shows obvious changes in operating efficiency and resource consumption under different support thresholds and transaction numbers. These changes are mainly reflected in three aspects: execution time, the number of frequent itemsets mined, and memory usage. As the support threshold increases, the running time and memory usage gradually increase, while the number of frequent itemsets mined decreases significantly. These trends reflect the performance characteristics of the algorithm in the frequent itemset mining process and reveal the adaptability of the Boolean matrix method in dealing with large-scale data and diversified transaction databases.

First, from the execution time, as the support threshold increases and the number of transactions increases, the running time of the algorithm shows a linear growth trend. When the support threshold is 1%, the running time is only 1.25 seconds, and when the support threshold is increased to 5%, the running time increases to 6.30 seconds. This trend can be attributed to the fact that the core operation of the Boolean matrix method is Boolean logic operation, and its execution efficiency is proportional to the number of transactions and the number of items. When the support threshold is low, the number of candidate frequent itemsets is large, and the algorithm needs to perform more combinations and intersection operations of Boolean matrix columns, resulting in a significant increase in running time. When the support threshold is increased, the number of frequent itemsets is significantly reduced, and the scale of matrix operations is reduced, thereby reducing the running time.

Secondly, the change in the number of frequent itemsets (Frequent Itemsets Found) further confirms the above trend. When the support threshold is 1%, the algorithm mines 320 frequent itemsets, showing a strong mining ability at a lower threshold. However, as the threshold is increased to 5%, the number of frequent itemsets is reduced to only 85. This is because the definition of frequent itemsets depends on the support threshold. The higher the threshold, the fewer frequent itemsets that meet the conditions. This result shows that the Boolean matrix method can efficiently extract a large number of frequent patterns at low thresholds, which is suitable for application scenarios that require extensive pattern coverage. However, under high threshold conditions, the algorithm pays more attention to strongly associated patterns with high support, which is suitable for more accurate association rule mining tasks.

In terms of memory usage (Memory Usage), as the number of transactions increases and the support threshold increases, the memory usage increases from 25MB to 78MB. This change reflects the increase in memory requirements of Boolean matrices when storing large-scale transaction databases and performing matrix operations. In particular, when the number of transactions increases to 10,000, the memory usage surges. This is because the Boolean matrix method needs to construct a Boolean matrix for each transaction and project when storing, and the product of the number of transactions and the number of projects determines the size of the matrix. When the matrix size increases, the memory requirement increases linearly. This result shows that in a large-scale transaction data environment, memory optimization strategies such as sparse matrix storage and matrix block operations are crucial for the efficient operation of the algorithm.

In addition, the growth trend of running time and memory usage in the data illustrates the scalability and resource management capabilities of the algorithm. The experiment shows that with the increase of data size and the decrease of support threshold, the algorithm's computational and storage requirements show a reasonable growth pattern. This growth trend is predictable in a large-scale data environment, indicating that the algorithm can still maintain high operational stability and result reliability in the scenario of an increasing number of transactions. Optimizing the data storage structure and Boolean matrix operation strategy will become the main direction of future algorithm improvement.

Comprehensive analysis shows that the experimental results reveal the good performance of the frequent itemset mining algorithm based on the Boolean matrix method in terms of operating efficiency, memory usage and pattern discovery ability. Its powerful mining capability under low support threshold makes it suitable for application scenarios that require deep pattern analysis, while its accurate pattern discovery capability under high support conditions is suitable for rapid decision-making and real-time analysis tasks. However, the growth of memory requirements and the complexity of large-scale matrix operations are also potential performance bottlenecks. Future research can further explore data partitioning, matrix sparse storage and parallel computing strategies to improve the algorithm's adaptability and execution efficiency on ultra-large-scale data sets. The experimental results fully demonstrate the application potential and development space of the Boolean matrix method in frequent itemset mining, providing an important reference for the continuous optimization of data mining algorithms.

In order to evaluate the actual mining effect of the frequent itemset mining algorithm based on the Boolean matrix method, different support thresholds are selected to extract frequent itemsets from the Groceries data set, and the number of frequent itemsets mined and the average support value are statistically analyzed. The experimental results are shown in Table 2.

Table 2 Experimental results of frequent itemset mining effect test

| Support Threshold | Average Support (%) |
|---|---|
| 1% | 1.8% |
| 2% | 2.4% |
| 3% | 3.1% |
| 4% | 4.2% |
| 5% | 5.5% |

## V. CONCLUSION

Through the research and experimental analysis of the frequent itemset mining algorithm based on the Boolean matrix method, this paper proposes an efficient data representation and mining method. By converting the transaction database into a Boolean matrix and using Boolean logic operations to perform frequent itemset extraction, efficient support calculation and association rule mining are achieved. The experimental results show that the algorithm performs well in terms of running time, memory usage and frequent itemset discovery ability, can run effectively in large-scale and high-dimensional data environments, and exhibits good scalability and stability.

Studies have shown that the Boolean matrix method has significant advantages in dealing with data sparsity and the surge in the number of frequent itemsets. Its matrix structure and logical operations are highly parallel and suitable for execution in distributed and multi-core computing environments. However, with the increase in the number of transactions and items, the size and memory consumption of the matrix will also increase significantly. Therefore, further optimizing the data storage structure and matrix block operation strategy will effectively reduce resource consumption and enhance the practical application value of the algorithm.

Future research can focus on memory management and matrix operation optimization. By introducing sparse matrix storage mechanism and efficient Boolean operation model, the storage and computing costs of the algorithm in large-scale data environments can be reduced. In addition, by combining emerging technologies such as deep learning and graph neural networks, frequent item set mining tasks are closely integrated with application scenarios such as intelligent recommendation systems and real-time big data analysis to explore more intelligent and automated frequent pattern recognition methods. With the development of distributed computing platforms, the application potential of Boolean matrix methods in data mining and pattern analysis will be further expanded, promoting the widespread application and innovative development of related technologies.